%% file: main_revised.tex
\documentclass[12pt]{iopart}

\usepackage[utf8]{inputenc}
\usepackage{graphicx}
\usepackage{dcolumn}
\usepackage{bm}
\usepackage{hyperref}
\begin{document}

\title[Estimating predictability of depinning dynamics by machine learning]{Estimating predictability of depinning dynamics by machine learning}

\author{Valtteri Haavisto, Marcin Mi{\'n}kowski, and Lasse Laurson}

\address{Computational Physics Laboratory, Tampere University, P.O. Box 600, FI-33014 Tampere, Finland}
\ead{valtteri.haavisto@tuni.fi}
\vspace{10pt}

\begin{abstract}
Predicting the future behaviour of complex systems exhibiting critical-like dynamics is often considered to be an intrinsically hard task. Here, we study the predictability of the depinning dynamics of elastic interfaces in random media driven by a slowly increasing external force, a paradigmatic complex system exhibiting critical avalanche dynamics linked to a continuous non-equilibrium depinning phase transition. To this end, we train a variety of machine learning models to infer the mapping from features of the initial relaxed line shape and the random pinning landscape to predict the sample-dependent staircase-like force-displacement curve that emerges from the depinning process. Even if for a given realization of the quenched random medium the dynamics are in principle deterministic, we find that there is an exponential decay of the predictability with the displacement of the line 
as it nears the depinning transition from below. Our analysis on how the related displacement scale depends on the system size and the dimensionality of the input descriptor reveals that the onset of the depinning phase transition gives rise to fundamental limits to predictability.
\end{abstract}

%
%
%
%
%

\section{Introduction}
From predicting tipping points in ecosystems by monitoring their loss of resilience~\cite{dai2012generic,scheffer2001catastrophic} to forecasting earthquakes by measuring subtle precursory seismic signals~\cite{bletery2023precursory}, the pursuit of forecasting the future behavior of complex systems~\cite{vlachas2022multiscale,craig2001bayesian} is a fascinating endeavor which, if successful, would be extremely useful. Yet, attempts of this kind have typically resulted in partial success only, and the prediction accuracy of, e.g., the occurrence time and magnitude of the next earthquake within a given region has not reached levels needed to make the predictions to be of much practical utility~\cite{kagan2000probabilistic}. While it is qualitatively rather obvious that complex systems are by nature difficult to predict due to the non-linear nature of their collective dynamics, what exactly is the fundamental limiting factor for their predictability in each case is often unclear.

A specific class of complex systems which are important in physics and materials science is given by elastic interfaces in quenched random media driven by an external force. They exhibit a non-equilibrium depinning phase transition between pinned and moving phases at a critical value of the external force~\cite{chauve2000creep, nattermann1992dynamics}, emerging from the interplay between quenched disorder of the medium, elasticity of the interface, and an external driving force. Examples include domain walls in ferromagnets~\cite{zapperi1998dynamics} and ferroelectrics~\cite{paruch2005domain}, contact lines in wetting~\cite{joanny1984model}, crack fronts in disordered solids~\cite{laurson2013evolution}, and dislocations in crystals~\cite{zapperi2001depinning}. Such systems belong to an even broader class of complex interacting systems driven out of equilibrium by external perturbations and exhibiting jerky avalanche dynamics as a response, ranging from earthquakes~\cite{richter1956magnitude} (as well as their lab-scale counterparts~\cite{laurenti2022deep}) to materials deformation~\cite{alava2014crackling,uchic2009plasticity,papanikolaou2017avalanches,budrikis2017universal,makinen2015avalanches,makinen2020propagating} and fracture~\cite{koivisto2016predicting,santucci2019avalanches} to neuronal~\cite{beggs2003neuronal,nandi2022scaling} and financial avalanches~\cite{biondo2013reducing}. Avalanches typically follow broad power-law size distributions and obey universal scaling laws~\cite{rosso2009avalanche}, analogously to scale-free features observed at continuous equilibrium phase transitions.

Due to the critical dynamics of the depinning phase transition, manifested as avalanches with a power-law size distribution characterized by power-law exponents that derive from the depinning transition critical point, predicting the time evolution of the interface in the proximity of the critical external force value is expected to be an inherently hard task~\cite{ramos2009avalanche}: Critical avalanches are by nature expected to always be on the edge of stopping while propagating. Therefore, for instance whether a given avalanche will become large or small is expected to be governed by small fluctuations that should be hard if not impossible to predict. Thus, it is typically implicitly assumed that a probabilistic description in terms of avalanche size distributions and related statistical quantities is, disregarding possible weak anticorrelations between successive events~\cite{le2020correlations}, the best possible description of such complex systems. 

On the other hand, for zero temperature and for a given realization of the quenched random medium as well as for a given initial condition and force ramp, the dynamics of the interface are deterministic within simple models such as the quenched Edwards-Wilkinson (qEW) equation~\cite{kardar1986dynamic,le2002two}, a paradigmatic model of driven elastic interfaces in random media we study in this work. Hence, given a (coarse-grained) description of the relaxed initial state of the interface as well as that of the quenched random medium, one might expect some degree of predictability of the ensuing depinning dynamics. The key questions then are how good this predictability is, and if perfect predictions cannot be established, what are the factors limiting it?

Here, we address these questions by training machine learning (ML) models to infer the mapping from features of the initial, relaxed line configuration and the quenched random medium to the force-displacement curve characterizing the transient approach towards the depinning transition from below. The quality of the predictions produced by the ML models (quantified by the coefficient of determination, $R^2$, between the predicted and the real values) is a useful measure of predictability. 
We note that in principle, a hypothetical ML model might exist which, when trained with enough data, would be able to perfectly predict the dynamics of the deterministic system we consider, resulting in $R^2=1$. However, our focus is in understanding how the $R^2$ of a given ML algorithm evolves with the displacement $d$ of the elastic line from the initial state. In other words, we do not claim that the ML models considered would give the upper limit of the predictability score, but rather argue that $R^2$ of a typical ML model, trained with a reasonable amount of data, evolves with $d$ in a certain way. As the ML model architecture is the same for each $d$, the only thing that changes for the different $d$ values is the displacement of the line itself. Hence, we argue that any evolution of $R^2$ with $d$ reflects changes in the properties of the system (its "predictability") rather than the properties of a specific ML model.
In this work, we employ different ML models ranging from linear regression to neural networks to convolutional neural networks, considering both one and two-dimensional (1D and 2D, respectively) input fields. This approach has parallels to recent studies that aim to predict the plastic deformation process of crystals subjected to applied stresses~\cite{salmenjoki2018machine,minkowski2022machine,minkowski2023predicting}, a complex problem which in the absence of static obstacles to dislocation motion (e.g., precipitates) is controlled by dislocation jamming~\cite{miguel2002dislocation}, resulting in an "extended critical" phase without a well-defined critical point~\cite{ispanovity2014avalanches,lehtinen2016glassy}. Here, the physical system we consider exhibits different dynamics that are controlled by the presence of a non-equilibrium phase transition critical point with well-known properties~\cite{kim2006depinning}, thus potentially allowing for a more clear-cut interpretation of the results concerning predictability. Indeed, a somewhat analogous problem is given by predicting the depinning dynamics of a dislocation pileup in a random quenched pinning potential~\cite{sarvilahti2020machine}, with the important difference that in the present case the elastic line is moving in a direction perpendicular to the average line orientation, thus always meeting new, previously unseen disorder as it propagates forward. Our study also complements a very recent study on the loss of memory of an elastic line driven through a disordered medium with biperiodic boundary conditions on its way to limit cycles~\cite{agoritsas2024loss}.

\begin{figure}[t!]
    \centering
    \includegraphics[width=0.75\columnwidth]{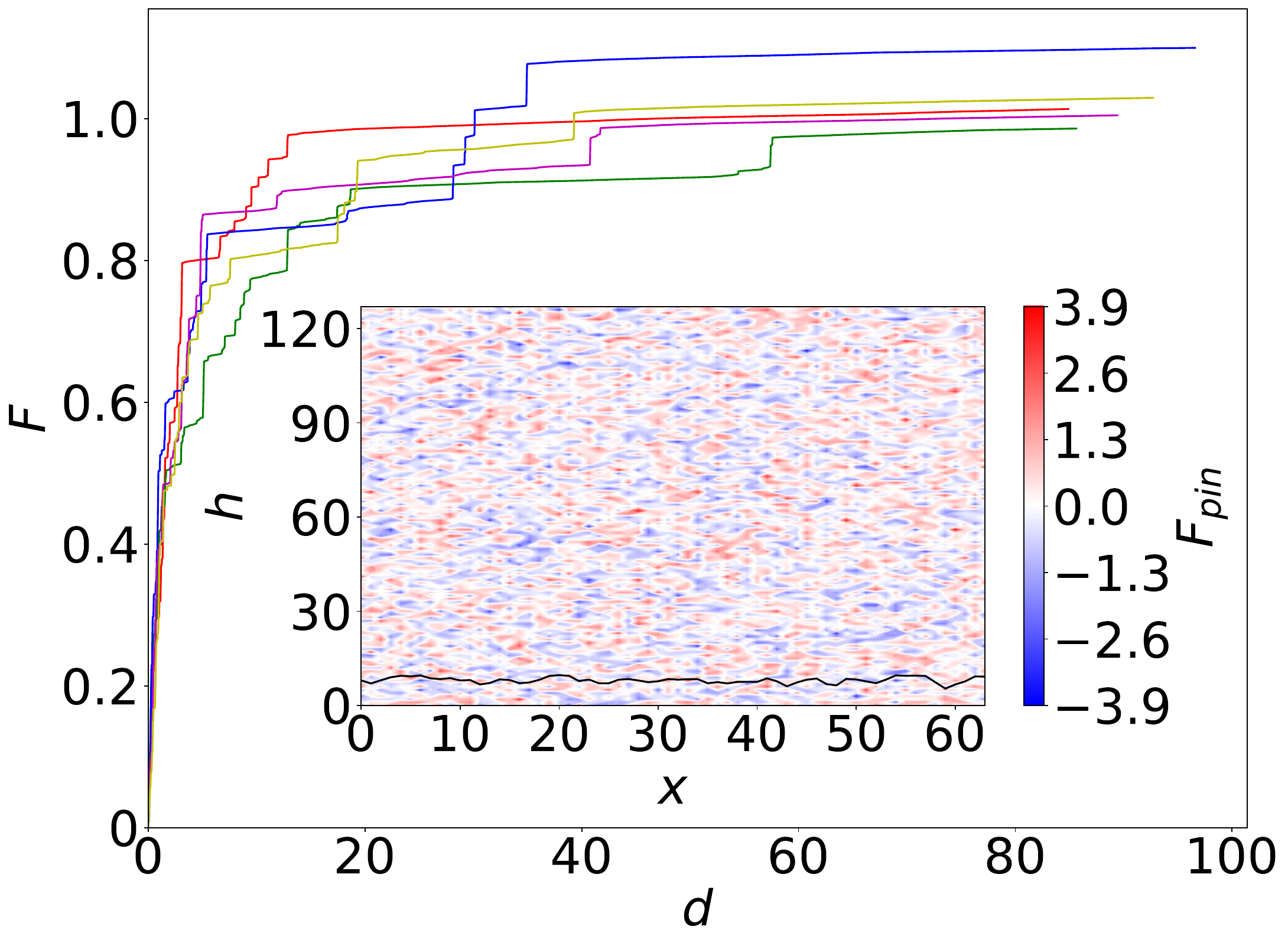}
    \caption{Main figure: Examples of force-displacement curves $F(d)$ for different realizations of the random pinning field $F_\textrm{pin}(x,h)$. Inset: An example of the relaxed line profile $h(x)$ (black line) and the corresponding quenched pinning field $F_\textrm{pin}(x,h)$ (colored according to the colorbar shown on the right). Our goal in this paper is to study to what extent the individual $F(d)$-curves can be predicted using information shown in the inset (initial line profile and the quenched pinning field).}
    \label{fig:1}
\end{figure}

Our results reveal an exponential decay of the predictability with the average interface displacement $d$ from the initial relaxed configuration, quantified by the coefficient of determination $R^2 \propto \exp (-d/d_0)$. The related displacement scale $d_0$ is a measure of how quickly the predictability decays with $d$. Possible interpretations of this could include that either the system loses its memory~\cite{keim2019memory} of the initial interface configuration as the interface moves forward and approaches the depinning phase transition from below, or that the proximity of the critical point is the main factor limiting predictability. We show that considering 1D input fields, characterizing the shape of the relaxed 1D interface and the pinning landscape only along it, leads to a smaller $d_0$ than considering 2D input fields including information of the quenched pinning field also ahead of the relaxed interface. Our main result is a size effect in the case of the 2D input fields showing that $d_0 \propto L^\zeta$, with $L$ the system size and $\zeta$ the roughness exponent, suggesting that since $d_0$ is proportional to the saturated interface width at criticality, the predictability of the interface dynamics is limited by the depinning phase transition the system is approaching from below. This is in contrast with the idea of the interface simply forgetting its initial state. Thus, our results provide a novel, quantitative perspective on how predictability of the dynamics of a complex system is limited by the proximity of a continuous non-equilibrium phase transition.

\section{Methods}
\label{sec:2}

\begin{figure*}[t!]
    \centering
    \includegraphics[width=\columnwidth]{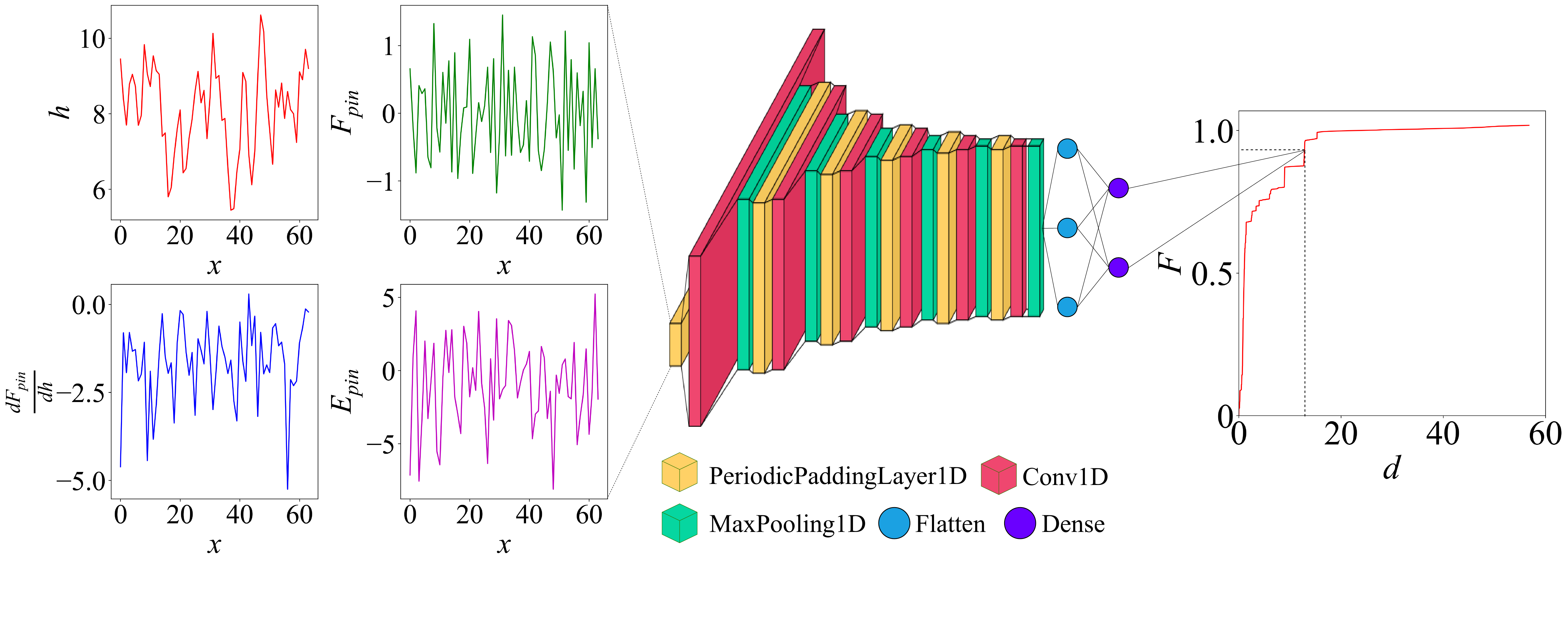}
    \caption{Schematic representation of the 1D CNN. The 1D fields along the relaxed line are fed into the CNN as the input arrays. That input is subsequently passed into padding, convolutional and pooling layers. The size of the array, represented here as its width, is reduced by half at each pooling layer until it is equal to 1. The number of parallel channels, represented by the height of the arrays, corresponds to the number of the convolutional filters. The flatten layer converts the final array of the width 1 into a linear array of the size equal to the number of filters. This is followed by a dense layer with a size equal to half the number of filters. Finally, another dense layer gives the prediction of $F(d)$ as the output. The procedure is then repeated for each $d$ to get a prediction for the entire $F(d)$ curve, i.e., we employ the "specialist" CNN approach expected to outperform a "generalist" model that would try to predict the entire $F(d)$ curve at once~\cite{sarvilahti2020machine}.}
    \label{fig:4}
\end{figure*}

\subsection{Quenched Edwards-Wilkinson equation}

To explore the predictability of the depinning dynamics of elastic interfaces in random media, we consider as an example system the qEW equation, defined by the equations of motion
\begin{equation}
\label{eq:1}
    \frac{\partial h_i}{\partial t} = \Gamma_0 \nabla^2 h_i + F_\textrm{pin}(i,h_i) + F
\end{equation}
for the continuous variables $h_i$, $i=1,...,L$ (with $L$ the system size), which constitute a discretized (along the average elastic line direction) description of the interface $h(x)$. $\Gamma_0$ is the stiffness of the interface, $F_\textrm{pin}$ is a position-dependent quenched random force (i.e., it is a function of $i$ and $h_i$ only), and $F$ is the external driving force. We employ periodic boundary conditions along the line (but not in the direction of motion perpendicular to the line direction), and discretize the Laplacian by setting $\nabla^2 h_i = h_{i+1} + h_{i-1} - 2h_i$. The quenched disorder field $F_\textrm{pin}(i,h_i)$ is constructed by first forming a regular discrete grid of size $L \times h_\mathrm{max}$ (with $h_\mathrm{max}>L$ depending on $L$; $h_\mathrm{max}=48$, 80, 128 and 192 for $L=16$, 32, 64 and 128, respectively) with unit spacing, and a random number is drawn from the standard normal distribution $N(0,1)$ to each grid point. Cubic spline interpolation is then employed separately for each $i$ to obtain a continuous disorder field along the direction of interface motion. An example of the resulting disorder field $F_\textrm{pin}$ is given in the inset of Fig.~\ref{fig:1}. In the simulations, Eqs.~(\ref{eq:1}) are integrated numerically using the Euler method, starting from a relaxation stage at $F=0$ from a flat profile at an initial height of $h_i = L/8$ for all $i$ (resulting in a somewhat rough line profile, see the black line in the inset of Fig.~\ref{fig:1} for an example), after which $F$ is ramped up from zero at a slow rate of $\partial F /\partial t = 0.0001$, chosen to approximate a quasistatically increasing $F$. The simulation is run until the first interface segment $i$ reaches a height close to but below the maximum allowed height of $h_i=h_\mathrm{max}$; this also roughly corresponds to the interface reaching the depinning transition at an effective $L$-dependent critical value of the external force, $F=F_c(L)$.

Unless stated otherwise, we consider a small system with $L=64$, which is expected to result in significant sample-to-sample variation in the response to external forces quantified by the force-displacement curve $F(d)$, where $d=\langle h_i(F)-h_i(0) \rangle_i$ is the $F$-dependent average interface displacement; our aim is to predict the sample-specific $F(d)$ using information of the initial configuration of the system, described by the relaxed line profile $h_i(0)$, $i=1,...,L$, and the quenched pinning landscape, as input. Examples of force-displacement curves $F(d)$ obtained for different realizations of the random pinning force are shown in the main figure of Fig.~\ref{fig:1}. One can observe the typical random-looking staircase-like shape of the curves, corresponding to a sequence of avalanches of interface propagation [the horizontal segments of $F(d)$, which tend to become larger upon increasing $F$] separated by the vertical segments corresponding to the external force increasing while the interface is pinned by the disorder. Due to the small system size considered, the response $F(d)$ exhibits rather pronounced sample-to-sample variations which we aim at predicting. 

To compute the database used to train the ML models, for $L=64$, the simulation procedure described above is repeated for 40000 different random realization of the pinning field $F_\textrm{pin}(i,h_i)$. The database then consists of the 40000 $F(d)$-curves and the corresponding descriptors of the initial relaxed configuration (see below for details on what these are for the different ML models). In order to address the question of how the system size $L$ affects predictability, we also compute three additional databases: 10000 configurations for system sizes $L=16$, 32 and 128.

\begin{figure*}[t!]
    \centering
    \includegraphics[width=\columnwidth]{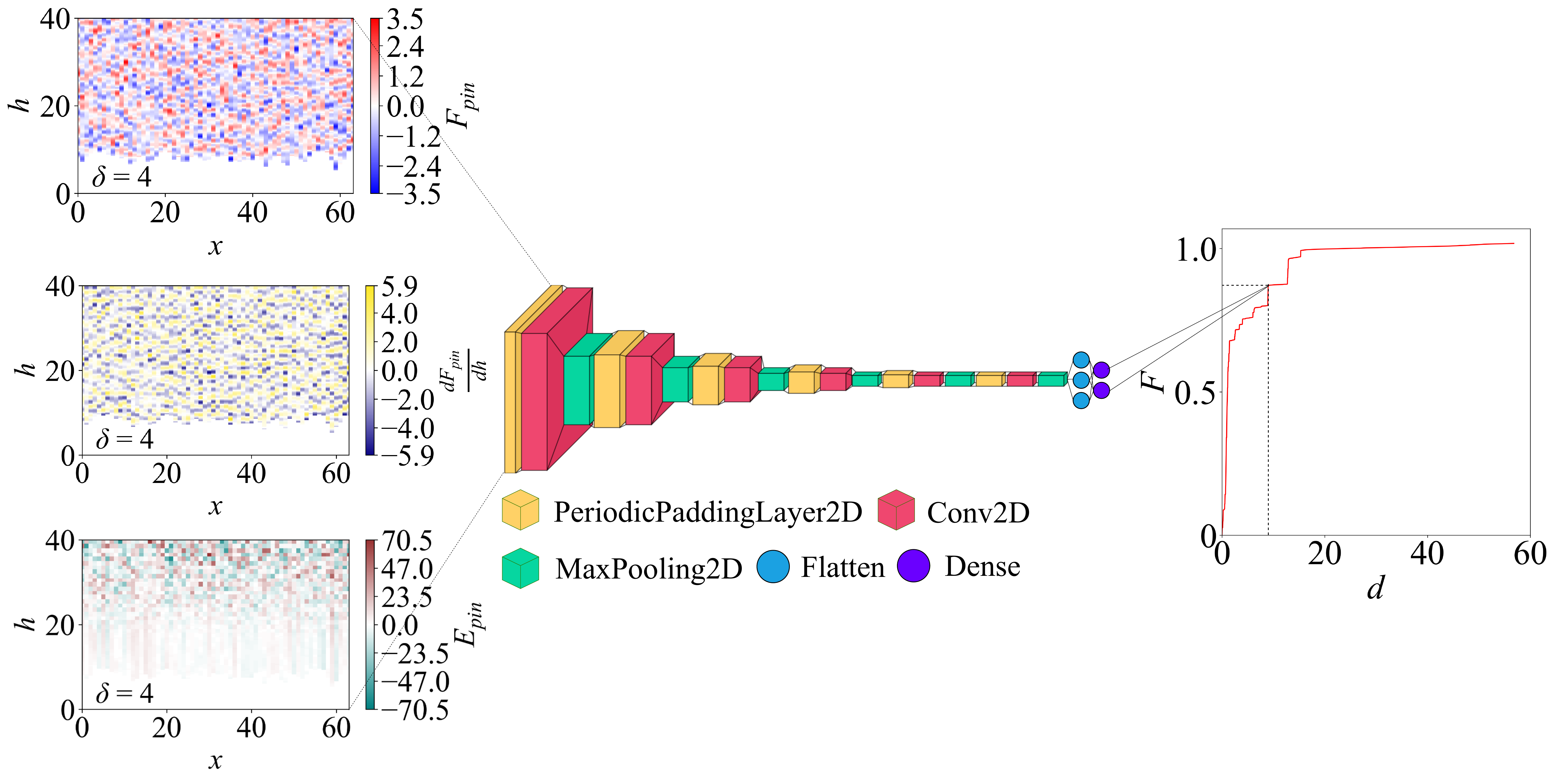}
    \caption{Schematic representation of the 2D CNN. The 2D fields $F_\mathrm{pin}$, $\mathrm{d}F_\mathrm{pin}/\mathrm{d}h$ and $E_\mathrm{pin}$ are fed into the CNN as the input arrays. That input is subsequently passed into padding, convolutional and pooling layers. The size of the array in the two dimensions, represented here as its width and height, is reduced by half at each pooling layer until it is equal to 1x1 if the array is square, or until the smaller of the dimensions is reduced to the size of 1 otherwise. The number of parallel channels, represented as the thickness of the arrays, corresponds to the number of the convolutional filters. The flatten layer converts the final array into a linear array of the size equal to the number of filters multiplied by the size of the final array in the longer direction (if the array is not square). This is followed by a dense layer with a size equal to half the number of filters. Finally, the last dense layer gives the prediction of $F(d)$ as the output. The procedure is then repeated for each $d$ to get a prediction for the entire $F(d)$ curve.}
    \label{fig:6}
\end{figure*}

\subsection{Machine learning models for predictability analysis}

To study the predictability of the depinning dynamics, we consider three different ML models: Linear regression (LR), fully connected neural networks (NNs) and convolutional neural networks (CNNs)~\cite{hastie2009elements}. Both NNs and CNNs are implemented in the Keras library of Python. Together these cover a broad range of different model types, ranging from linear mappings from user-defined features (LR, see Appendix A) to non-linear mappings without the need for feature engineering by the user (CNN). For LR and NN, the input features used are statistical properties (average, standard deviation, kurtosis, skewness, maximum, and the absolute values of the first and second Fourier coefficients along the interface) of the relaxed line profile $h(x)$, as well as of three 1D profiles measured along the relaxed line profile: the pinning force $F_\mathrm{pin}(x)$, its derivative $\mathrm{d}F_\mathrm{pin}/\mathrm{d}h(x)$, and the pinning energy $E_\mathrm{pin}(x)=-\int_0^{h(x)} F_\mathrm{pin}\mathrm{d}h$, see Fig.~\ref{fig:4} and Appendix A for example profiles. 

For the CNN, the input consists of 1D or 2D fields without feature engineering: The 1D fields are the ones from which the statistical features used for LR and NN are extracted (see again Fig.~\ref{fig:4} and Appendix A). The 2D fields include information of the pinning field also ahead of the relaxed line profile, i.e., for $h>h(x)$, where $h(x)$ is the relaxed line profile; they are given by $F_\mathrm{pin}(x,h)$, $\mathrm{d}F_\mathrm{pin}/\mathrm{d}h(x,h)$ and $E_\mathrm{pin}(x,h)$, see Fig.~\ref{fig:6} and Appendix A for examples. The 2D fields contain zeros below the relaxed line profile, i.e., for $h<h(x)$, and hence the interface between zeros and (mostly) non-zero values contains information of the relaxed line configuration; we note that no separate 1D initial line profile is fed to the 2D CNN, and hence the task of figuring out the initial $h(x)$ from the interface between zeros and non-zero values is left to the CNN. Thus, overall, the 1D and 2D fields contain information of same kinds of quantities, but the 2D fields (which are given with different resolutions $\delta$) include more information, since they include knowledge of the pinning field also ahead of the relaxed line profile. The resolution of these fields is such that there are always $L$ pixels in the $x$-direction, and $\delta$ (where $\delta$ defines the resolution) pixels per unit displacement in the $h$-direction. For details on the ML models and the input features used, see Appendix A.

\section{Results}
\label{sec:3}

\subsection{Feature importance}

\begin{figure*}[t!]
    \centering
    \includegraphics[width=\columnwidth]{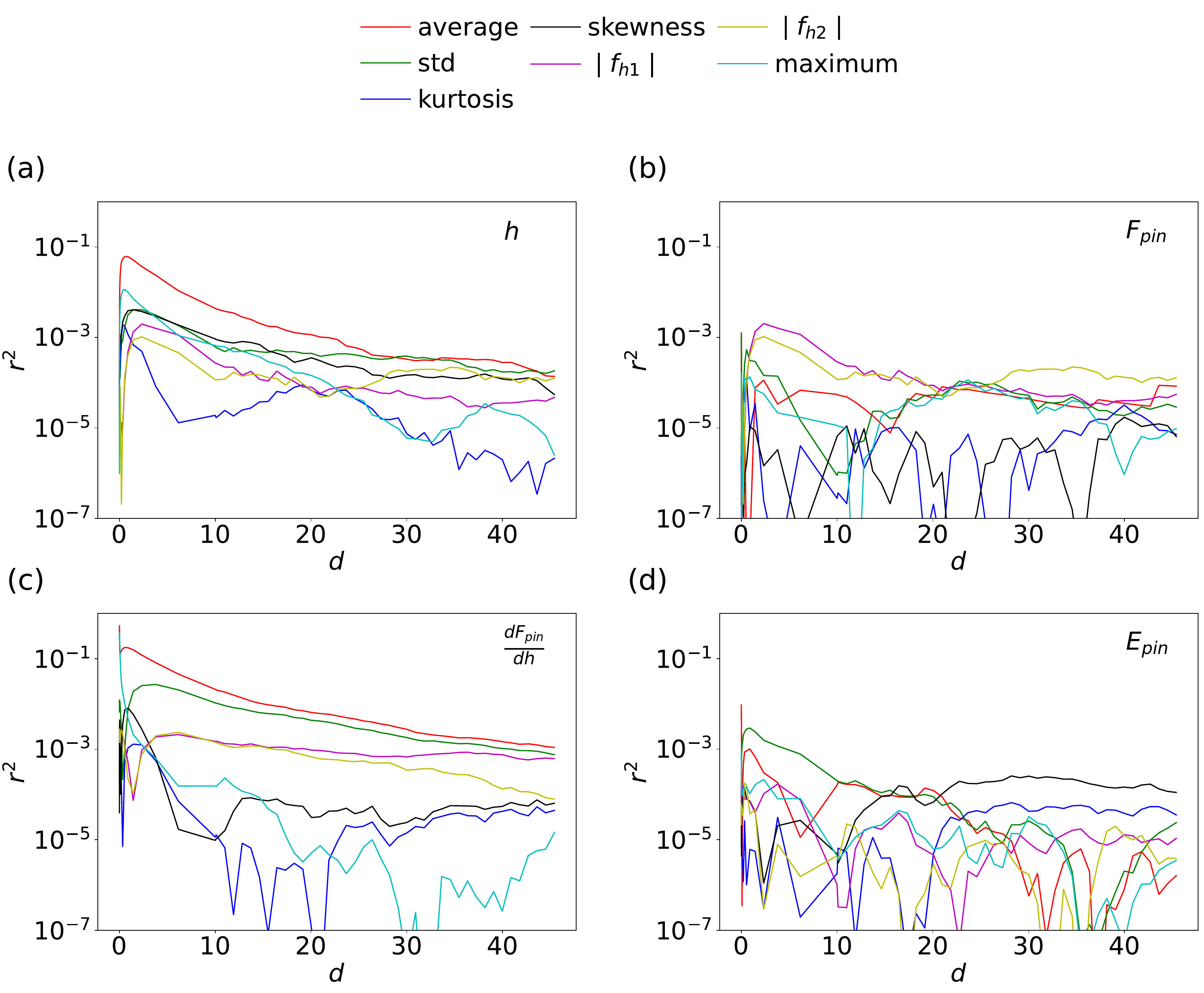}
    \caption{The square of the sample correlation coefficient $r^2$ of the different statistical features of the fields fed into the LR and NN models with the external driving force $F(d)$ as the function of the displacement $d$. (a) The relaxed interface profile $h$, (b) the pinning field $F_\mathrm{pin}$, (c) the pinning force derivative $\mathrm{d}F_\mathrm{pin}/\mathrm{d}h$, and (d) the pinning energy $E_\mathrm{pin}$.}
    \label{fig:7}
\end{figure*}

In order to shed some light on which features fed to the LR and NN models are important in determining the external force value $F(d)$ for a given displacement $d$, we start by considering the linear correlations (quantified by the square of the sample correlation coefficient, $r^2$) between the values of the various descriptors and $F(d)$. These are shown in Fig.~\ref{fig:7}, considering separately the features extracted from $h$, $F_\mathrm{pin}, \mathrm{d}F_\mathrm{pin}/\mathrm{d}h$, and  $E_\mathrm{pin}=-\int_0^{h_i} F_\mathrm{pin}\mathrm{d}h$, respectively. The general observation is that most of the features are very weakly correlated with $F(d)$, and the $r^2$-values for those features that exhibit slightly stronger correlations clearly decay with $d$. One may note that especially for small $d$ the strongest correlations are found for the average values of $\mathrm{d}F_\mathrm{pin}/\mathrm{d}h$ and $h$. The former relates to the stability of the relaxed interface configuration, and the latter quantifies the drift of the interface position away from its initial position during relaxation. However, none of the features alone exhibit strong correlations, and hence we proceed to consider their combined effect by employing the ML models.

\subsection{Predictability: 1D fields}

To study the overall predictability when all the different features (either 1D or 2D) are used together as input, the ML models introduced above are considered. The predictability is quantified by the coefficient of determination $R^2$, defined by
\begin{equation}
R^2=1-\frac{\sum_{i=1}^{N}[F_i^\mathrm{true}(d)-F_i^\mathrm{pred}(d)]^2}{\sum_{i=1}^{N}[F_i^\mathrm{true}(d)-\langle F_i^\mathrm{true}(d) \rangle]^2},
\end{equation}
where $F_i^\mathrm{true}(d)$ is the true value [i.e., the value obtained from the numerical solution of Eq.~(\ref{eq:1})] of the force $F(d)$ at a specific displacement $d$ for the $i$th sample, $\langle F_i^\mathrm{true}(d) \rangle$ is its mean value, $F_i^\mathrm{pred}(d)$ is the value predicted by the ML model, and $N$ is the total number of samples in the set. $R^2=1$ would imply perfect predictability, while $R^2=0$ would mean that the model would simply predict the average response for all samples.

\begin{figure}[t!]
    \centering
    \includegraphics[width=0.75\columnwidth]{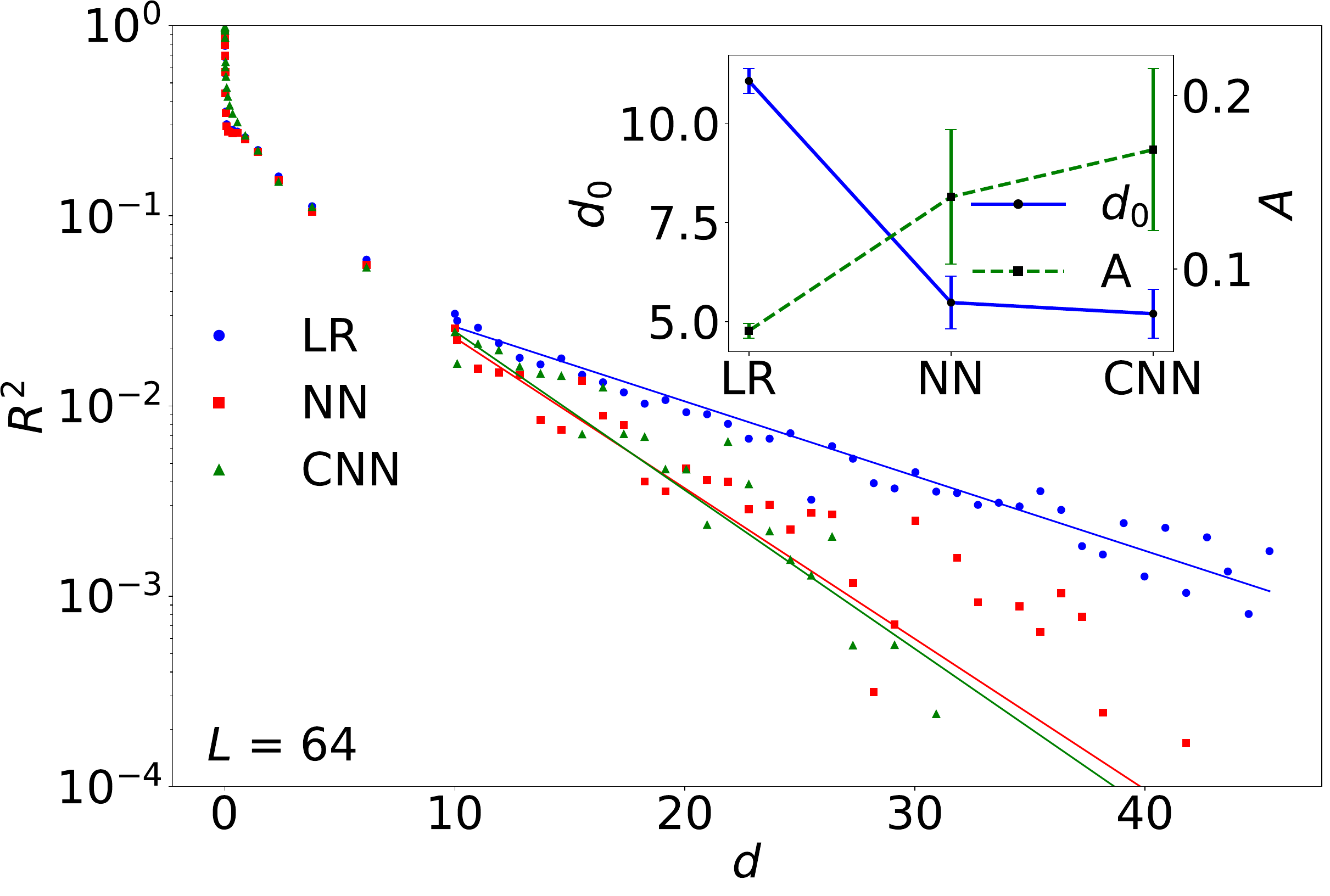}
    \caption{The coefficient of determination $R^2$ for the three ML models with the 1D input fields as a function of displacement $d$. Solid lines are exponential fits (Eq. \ref{eq:5}). The inset shows the corresponding fitting parameters $d_0$ and $A$.}
    \label{fig:8}
\end{figure}

We start by considering predictability using the 1D input fields. We find (see Fig.~\ref{fig:8} showing the test set $R^2$ for LR, NN and 1D CNN) that for $d$ larger than $\sim 10$ (after the small-$d$ peak for which $R^2$ is not much below 1), $R^2$ decays roughly exponentially with the displacement $d$, 
\begin{equation}
    \label{eq:5}
    R^2(d) = A\mathrm{e}^{-\left(\frac{d}{d_0}\right)},
\end{equation}
where $d_0$ is the characteristic displacement scale of the exponential decay, quantifying how quickly information of the initial configuration is lost (or forgotten) by our ML models with increasing $d$. For LR with $L1$ regularization, we obtain $d_0 \approx 11$ (inset of Fig.~\ref{fig:8}). Perhaps surprisingly, NN and 1D CNN result in worse predictability for large $d$ with smaller $d_0$ of the exponential decay ($d_0 \approx 6$ and $d_0 \approx 5$ for NN and 1D CNN, respectively, see again the inset of Fig.~\ref{fig:8}). This suggests that the predictability is limited by the information content of the 1D descriptors so that increasing the complexity of the ML model (by moving from LR to NN and 1D CNN) does not improve the result but rather makes it worse for the finite training database at hand. On the other hand, one may notice that for $d \rightarrow 0$, $R^2 \approx 1$, suggesting that the detailed knowledge provided here to the ML models of the initial configuration contains enough information for very good predictability of the "linear response" away from the initial configuration.

\begin{figure}[t!]
    \centering
    \includegraphics[width=0.55\columnwidth]{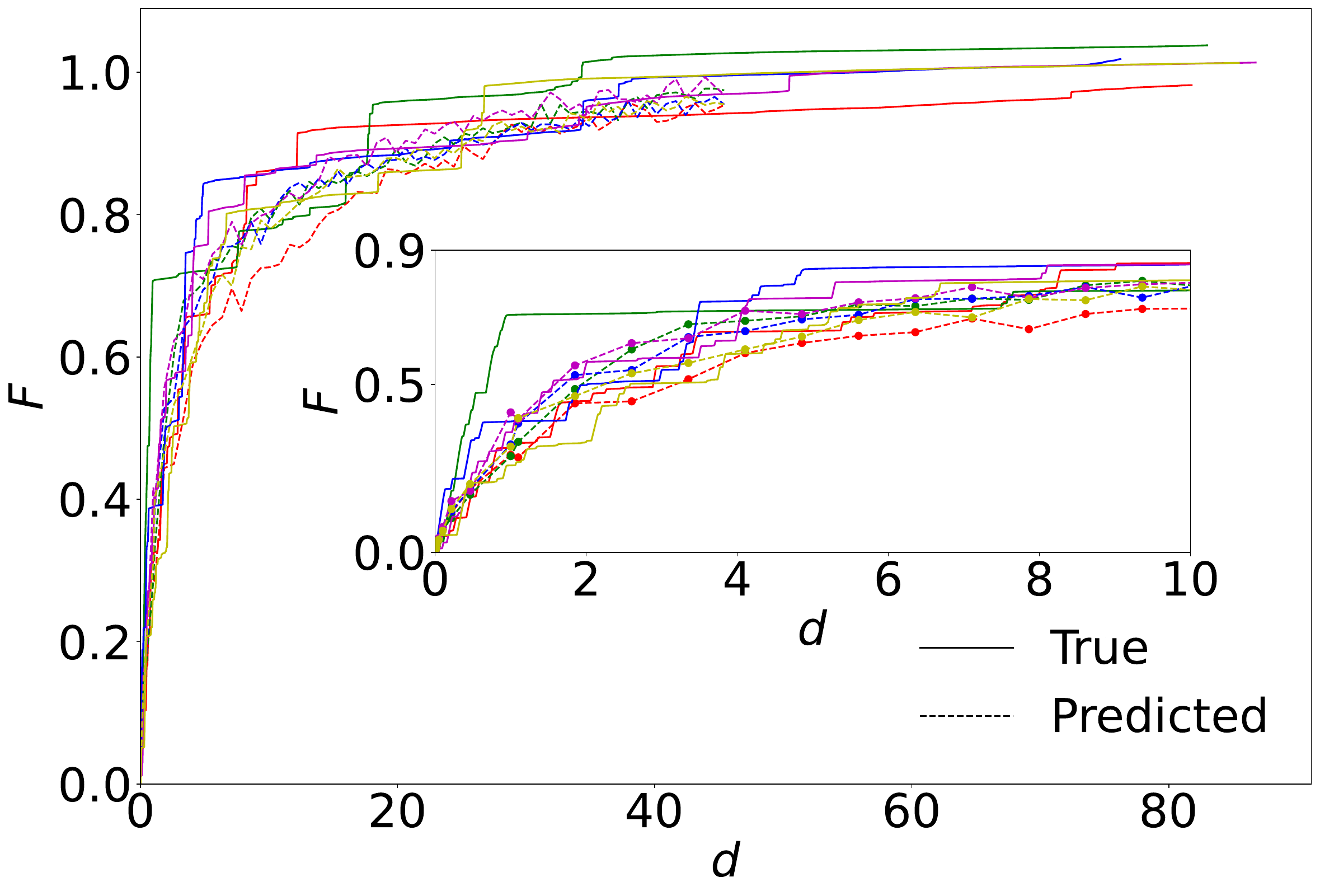}
    \caption{5 examples of actual force-displacement curves $F(d)$ (solid lines) and their predictions (dashed lines) by the 2D CNN for $L = 64$ and $\delta = 1$. The inset is zoomed to the lower displacements to more clearly illustrate the differences between the true and predicted $F(d)$ curves, where the dots are the points where force values were predicted.}
    \label{fig:true_pred_L64}
\end{figure}

\begin{figure}[t!]
    \centering
    \includegraphics[width=\columnwidth]{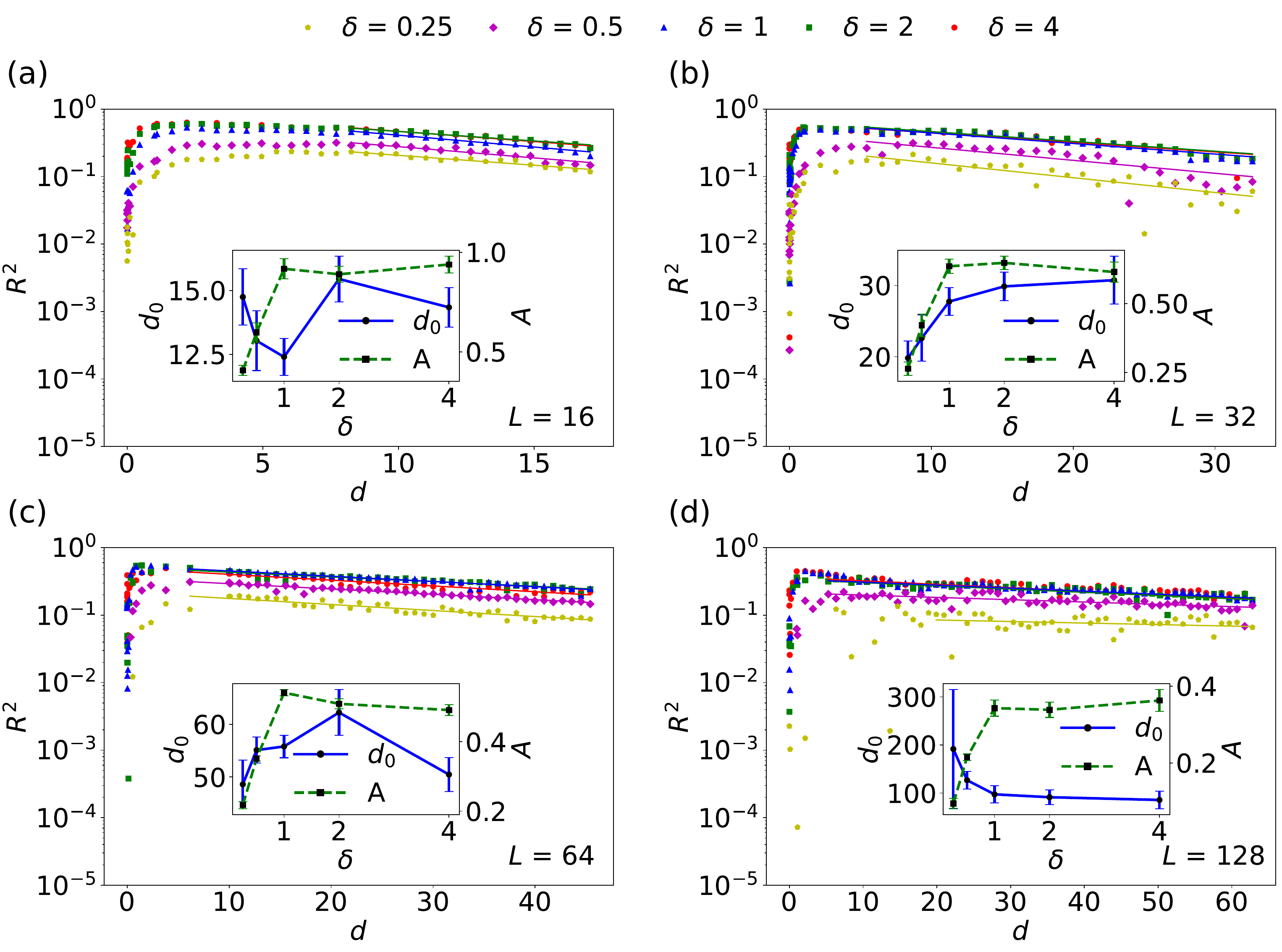}
    \caption{The coefficient of determination $R^2$ for the 2D CNN as a function of displacement $d$ for different system sizes $L$ [with $L=16$, 32, 64 and 128 shown in (a), (b), (c) and (d), respectively] and resolutions of the input $\delta$ (symbols). Lines are exponential fits [Eq.~(\ref{eq:5})]. The insets show the corresponding fitting parameters $d_0$ and $A$.}
    \label{fig:9}
\end{figure}

\subsection{Predictability: 2D fields}
As found above, the 1D descriptors do not contain enough information for good predictability of $F(d)$ for large $d$. This is to be expected as no information of the quenched pinning field above the relaxed interface configuration, something that to a large extent will determine the interface dynamics for $d>0$, is included in them. This effect could in principle be quantified further by randomizing the pinning forces ahead of the relaxed interface, resulting in an ensemble of future interfaces (and corresponding avalanches). The variance of this ensemble would be a measure of the randomness induced by ignoring the values of the future pinning forces. To improve from this, we consider here the 2D CNN which is fed a 2D representation of the pinning field $F_\textrm{pin}$ ahead of the relaxed interface, together with the corresponding 2D fields of $E_\mathrm{pin}$ and $\mathrm{d} F_\mathrm{pin}/\mathrm{d}h$ (see again Fig.~\ref{fig:6} and Appendix A). 

\begin{figure}[t!]
    \centering
    \includegraphics[width=0.55\columnwidth]{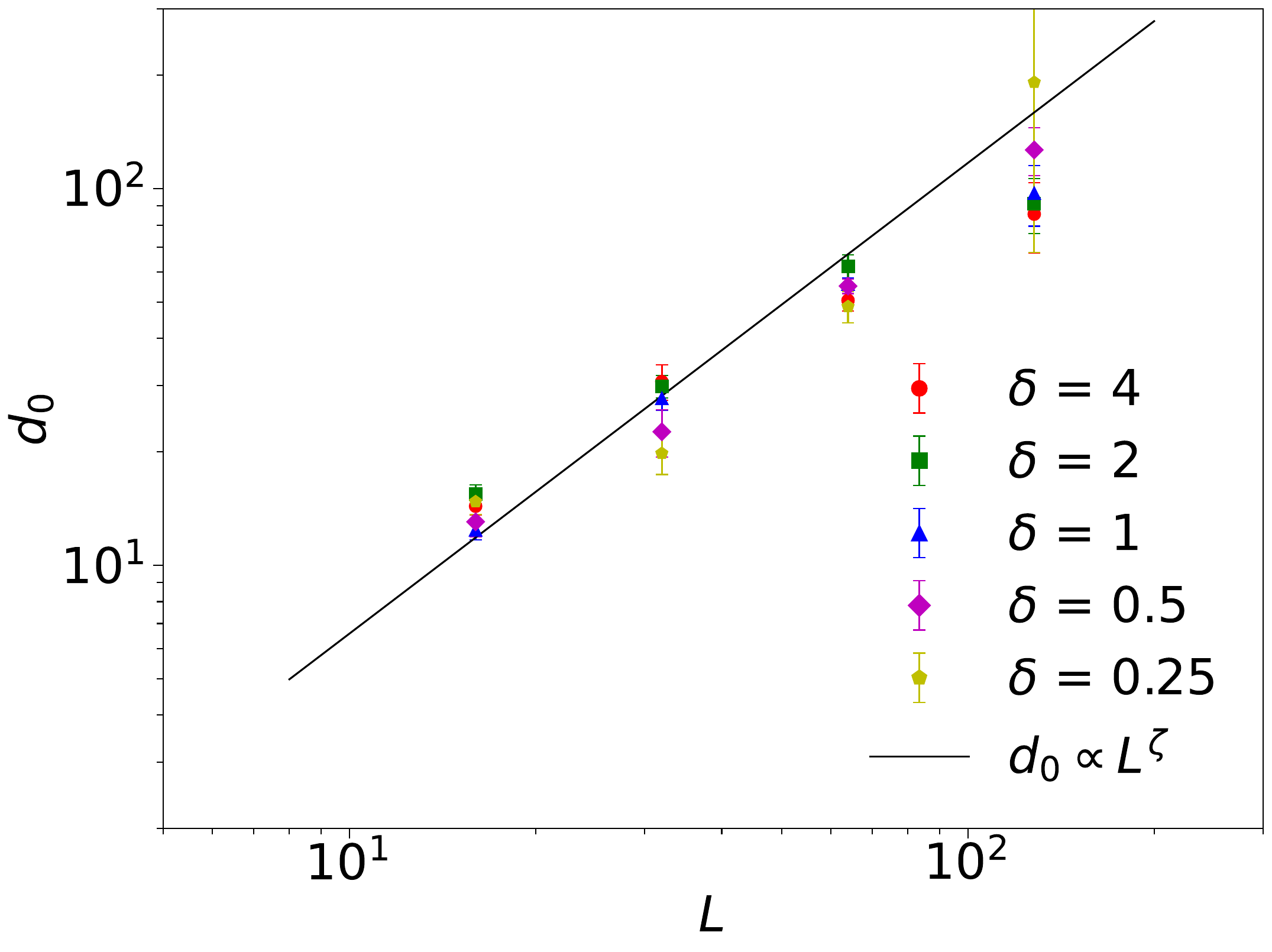}
    \caption{$d_0$ for the 2D CNN (from Fig.~\ref{fig:9}) as the function of the system size $L$ for different resolutions $\delta$ (symbols). Line is a fit of the form of $d_0 \propto L^{1.25}$ [Eq.~(\ref{eq:size_e})].}
    \label{fig:10}
\end{figure}

We start by showing in Fig.~\ref{fig:true_pred_L64} 5 randomly selected examples of both the actual force-displacement curves $F(d)$ and their predictions by the 2D CNN for $L = 64$ and $\delta = 1$. One can see that the 2D CNN is poor in predicting individual avalanches, but it predicts the overall trend of the $F(d)$ curves. This suggests that critical-like avalanches are the limiting factor for predictability.
Fig.~\ref{fig:9} then shows the resulting test set $R^2$ values as a function of $d$ for different resolutions $\delta$ of the pinning field descriptors, now considering four different system sizes $L$ to address the question of possible size effects in  predictability. Again, an exponential decay of $R^2$ with $d$ [Eq.~(\ref{eq:5})] is observed for large enough $d$. We note that now $d_0$ is significantly larger than in the case of the 1D fields, varying between $d_0 \approx 12$ and 15 for $L=16$, $d_0 \approx 20$ and 30 for $L=32$, $d_0 \approx 50$ and 60 for $L=64$, and $d_0 \approx 100$ and 200 for $L=128$, respectively, for the different resolutions $\delta$ considered (see the insets of Fig.~\ref{fig:9}). Moreover, the exponential decay starts from a larger $R^2$-value for small $d$ than in the case of the 1D input fields, and hence the large-$d$ predictability (while not great) is now much better than when considering 1D input fields. One may also notice that both $d_0$ and the prefactor $A$ depend on $\delta$ (see again the inset of Fig.~\ref{fig:9}), such that overall the best predictability is obtained for $\delta = 1$ or higher. This makes sense as $\delta=1$ matches the resolution (or density) of discrete random numbers from which the continuous $F_\textrm{pin}$ field is obtained via spline interpolation. Thus, for $\delta=1$ all the information of $F_\textrm{pin}$ is contained in the descriptor, but at the same time the number of pixels is not unnecessarily large so that the 2D CNN remains not too complex, something that helps to avoid overfitting. One may also note that the maximum of $R^2$ for small $d$ is now significantly below 1, in contrast to the $d \rightarrow 0$ value of $R^2 \approx 1$ found above for the 1D ML models, and in the $d \rightarrow 0$ limit $R^2$ actually tends to 0. This is because here the initial configuration is characterized  with a finite resolution $\delta$, and hence good predictions can be made only for displacement values exceeding the "pixel size" of the 2D input fields.

The above results show that for the 2D input fields, the CNN predictability and especially the $d_0$-value depends on $L$, such that the exponential decay of $R^2$ with $d$ is slower for larger $L$. Fig.~\ref{fig:10} shows the $d_0$-values for different $\delta$ as a function of $L$. The data seems to be, within error bars, consistent with the scaling with the roughness exponent $\zeta \approx 1.25$~\cite{rosso2007numerical,kim2006depinning}, i.e., 
\begin{equation}
\label{eq:size_e}
d_0 \propto L^{\zeta}. 
\end{equation}
When approaching the depinning transition from below starting from a flat initial configuration, one expects that the growing roughness $W$ of the interface is linked to its average height (here, the interface displacement $d$) via $W \propto d \propto \xi_x^{\zeta}$ (with $\xi_x$ the correlation length along the interface), so that the displacement $d^*$ at which the saturated roughness is reached at the depinning critical point (as $\xi_x \rightarrow L$) scales as $d^* \propto L^{\zeta}$. Thus, the observation that $d_0 \propto L^{\zeta} \propto d^*$ [Eq.~(\ref{eq:size_e})] suggests that displacements $d$ up to which meaningful prediction can be made (quantified by the $d_0$-value) are limited by the displacement corresponding to the saturated roughness at the finite-$L$ depinning threshold. In other words, predictability is limited by the depinning phase transition from a pinned to a moving phase the system is approaching from the pinned phase. Interestingly, this scaling would indicate that since $\zeta > 1$, $d_0/L \propto L^{\zeta-1}$ would diverge in the thermodynamic limit $L \rightarrow \infty$ for the qEW equation studied here. One may note that the prefactor $A$ has the opposite trend with $L$ (see again the insets of Fig.~\ref{fig:9}), possibly because larger $L$ implies a more complex CNN with more trainable parameters negatively affecting the learning for a database of finite size. Notice however that the $L$ dependence of $A$ is much weaker than that of $d_0$: in the range of $L$-values considered, $A$ changes by a factor of $\sim 2$, while $d_0$ changes by almost an order of magnitude. Assuming that these trends persist for a broader range of $L$-values, this would imply that for a given large enough displacement $d$, $R^2$ would be larger in a system with a larger $L$. However, one should notice that the sample-to-sample fluctuations of $F(d)$ are expected to vanish in the thermodynamic limit, and hence predicting the sample-specific $F(d)$, as done here, only makes sense for a finite $L$.

\begin{figure}[t!]
    \centering
    \includegraphics[width=0.7\columnwidth]{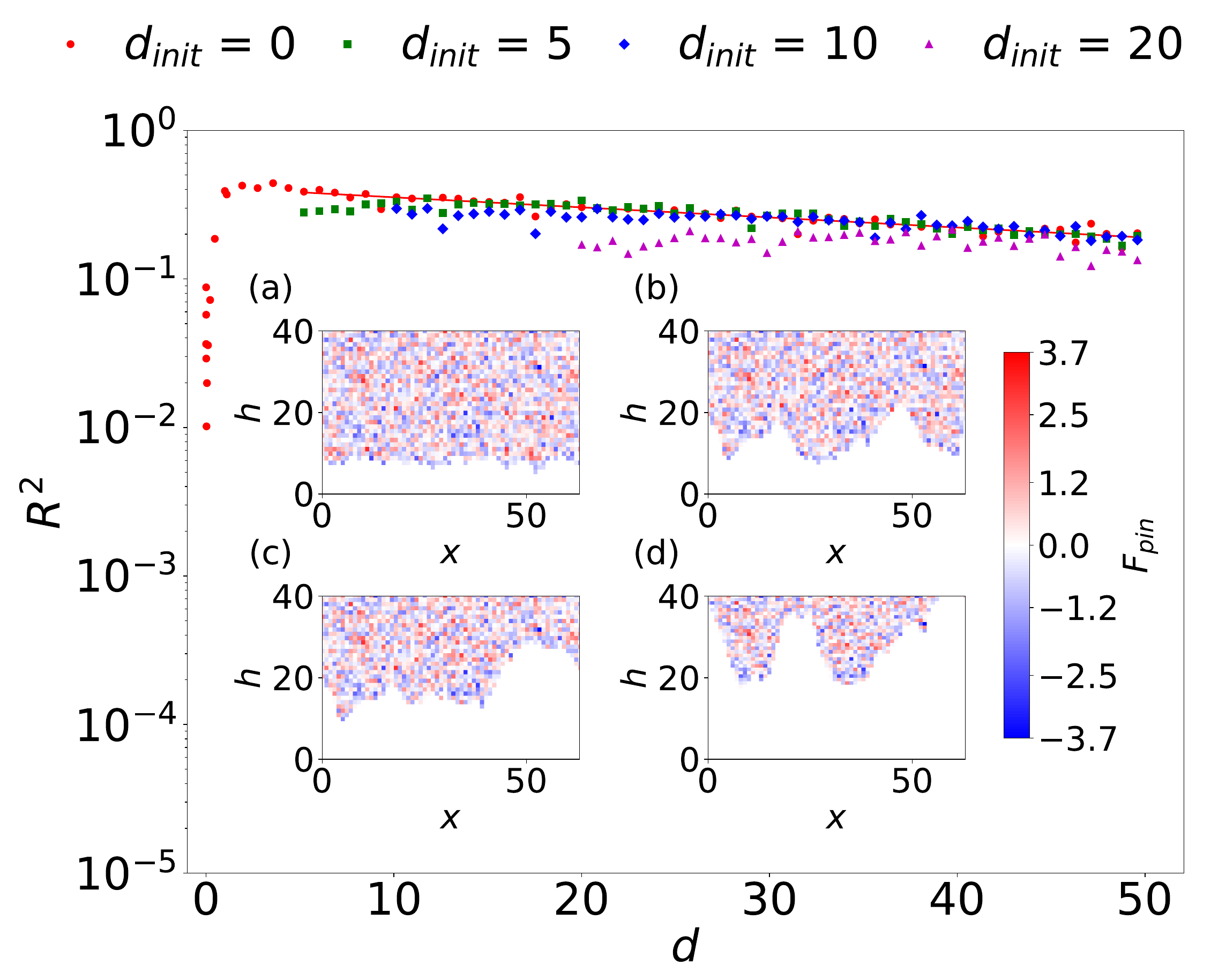}
    \caption{Main figure: $R^2(d)$ as a function of $d$ for the 2D CNN trained with input configurations obtained at different initial displacements $d_{init}$, considering $L=64$, $\delta = 1$ as an example system. The dataset is different from the one shown in Fig.~\ref{fig:9}, and consists of 10000 configurations. Fitting parameters corresponding to the red solid line for $d_{init}=0$ are $A\approx0.41$ and $d_0\approx64.1$. Inset: Examples of the pinning force landscapes used as input for the 2D CNN. (a) $d_{init}=0$, (b) $d_{init}=5$, (c) $d_{init}=10$, and (d) $d_{init}=20$. As before, the interface between white (zero) and colorful (non-zero) regions corresponds to the line profile.}
    \label{fig:d_init_L64_res_1}
\end{figure}

Finally, we consider an alternative approach to analyze the underlying reason behind the loss of predictability 
as $d$ increases. Specifically, in addition to using the relaxed initial state as input, we consider configurations 
extracted at different initial displacements $d_{init}>0$ [corresponding to different non-zero average $F(d_{init})$-values] 
and extract the 2D input fields from those; see the inset of Fig.~\ref{fig:d_init_L64_res_1} for examples 
corresponding to $d_{init}=0, 5, 10$ and 20. The 2D CNN is then trained to learn the $F(d)$ curve for 
$d>d_{init}$ by using the 2D input fields extracted at the different $d_{init}$-values considered. The main figure 
of Fig.~\ref{fig:d_init_L64_res_1} shows the resulting $R^2$'s as a function of $d$ for $d>d_{init}$, 
considering different $d_{init}$-values. The key observation is that after an initial transient, the curves for 
$d_{init}>0$ tend towards the $d_{init}=0$ curve from below, such that $R^2(d)$ for $d_{init}>0$ does not seem to 
exceed that obtained for $d_{init}=0$. This indicates that by providing the CNN with more information on how the 
dynamics has progressed up to $d=d_{init}$ (by feeding it the line configuration at $d=d_{init}>0$), one cannot 
improve the predictions from those of $d_{init}=0$. Hence, this analysis further supports the idea that the proximity 
of the critical point is the main predictability-limiting factor. Specifically, it does not seem like the decrease of 
predictability with increasing $d$ could be attributed to the system 
"forgetting its initial state"~\cite{agoritsas2024loss,keim2019memory}. Rather, the main reason for reduced 
predictability appears to be the avalanche dynamics becoming more critical-like as the depinning transition is 
approached from below.

\section{Discussion and conclusions}
\label{sec:4}
To conclude, we have found an exponential decay with the displacement $d$ of the predictability (quantified by the coefficient of determination, $R^2$) of the depinning dynamics of elastic interfaces in quenched random media. We stress again that this is not meant to imply that the best possible ML model trained with an arbitrarily large database could not reach a larger $R^2$. Rather, we fix the ML model and database size, and observe how the predictive ability of the model evolves with $d$, and expect such evolution to reflect properties of the physical system considered rather than those of the ML model. The amplitude $A$ and the characteristic scale $d_0$ of the exponential decay of $R^2$ with $d$ are found to depend on whether the input field is 1D or 2D. The observation that the best result is obtained by using the 2D CNN highlights the crucial role of features of the sample-specific pinning field ahead of the relaxed interface configuration, in addition to the relaxed line profile. The exponential decay of $R^2(d)$ can be linked to the critical-like avalanche dynamics that becomes more pronounced with increasing $d$ as the system is approaching the depinning phase transition. 
This behaviour is in contrast with findings from analogous studies of other complex avalanching systems including 2D discrete dislocation dynamics (DDD)~\cite{salmenjoki2018machine} and dislocation pileups~\cite{sarvilahti2020machine}, where a non-monotonic dependence of $R^2$ on strain has been found. This is likely due to the presence of features that are conserved during the dynamics in those systems (the 1st Fourier coefficient of the geometrically necessary dislocation density in the direction perpendicular to the dislocation motion in 2D DDD~\cite{salmenjoki2018machine}, and the repeated sampling of the same pinning field by the dislocation pileup with periodic boundary conditions~\cite{sarvilahti2020machine}). Both above-mentioned properties are absent in the present case.  

Our main result, obtained by studying the size dependence of predictability and supported by consireding input fields at various nonzero initial displacements $d_{init}$, reveals that the onset of critical dynamics due to the depinning phase transition the system is approaching from below is the limiting factor for the interface displacements up to which meaningful predictions of its dynamics can be made. It is important to emphasize that this limitation to predictability emerges even in cases such as ours where the interface dynamics, for a given realization of the quenched random medium, is governed by deterministic equations of motion. This provides an interesting perspective to understand the limits of predictability in a broad range of complex systems exhibiting a continuous non-equilibrium phase transition. For the specific case at hand, i.e., the qEW equation, the observed scaling $d_0 \propto L^{\zeta}$ together with the super-rough nature of the interface ($\zeta \approx 1.25 > 1$) implies that $d_0/L$ increases with $L$. For other elastic interfaces (such as the long-range elastic string~\cite{laurson2010avalanches,laurson2013evolution}) with $\zeta<1$, we hence expect $d_0/L$ to decrease with $L$, suggesting that predicting the large-displacement dynamics of long interfaces would be harder than in the present case. Indeed, it would be interesting to extend the present study by testing our ideas in the context of depinning of driven elastic interfaces in random media in higher dimensions (2D elastic membranes in 3D quenched random media~\cite{zapperi1998dynamics}) and with other types of elasticity, including long-range and non-linear elasticity (i.e., the Kardar-Parisi-Zhang equation~\cite{kardar1986dynamic} with quenched noise). Moreover, here the initial states have been prepared by relaxation from a flat configuration at $F=0$ (as well as by considering finite initial displacements $d_{init}$), but one could also create them by first quasistatically ramping up the force followed by a relaxation at zero force (in analogy with the pre-strained dislocation configurations studied in Ref.~\cite{salmenjoki2018machine}), thus allowing one to study possible history dependence of predictability. 

\section{Acknowledgments}
LL thanks Mikko Alava for interesting discussions. The authors acknowledge the computational resources provided by Tampere Center for Scientific Computing (TCSC), and support of the Research Council of Finland via the Academy Project COPLAST (Project no. 322405).

\section*{Appendix A: Details of the machine learning models}

\renewcommand{\thesubsection}{\arabic{subsection}}

\subsection{Linear regression}

\begin{figure*}[ht!]
    \centering
    \includegraphics[width=0.45\pdfpagewidth]{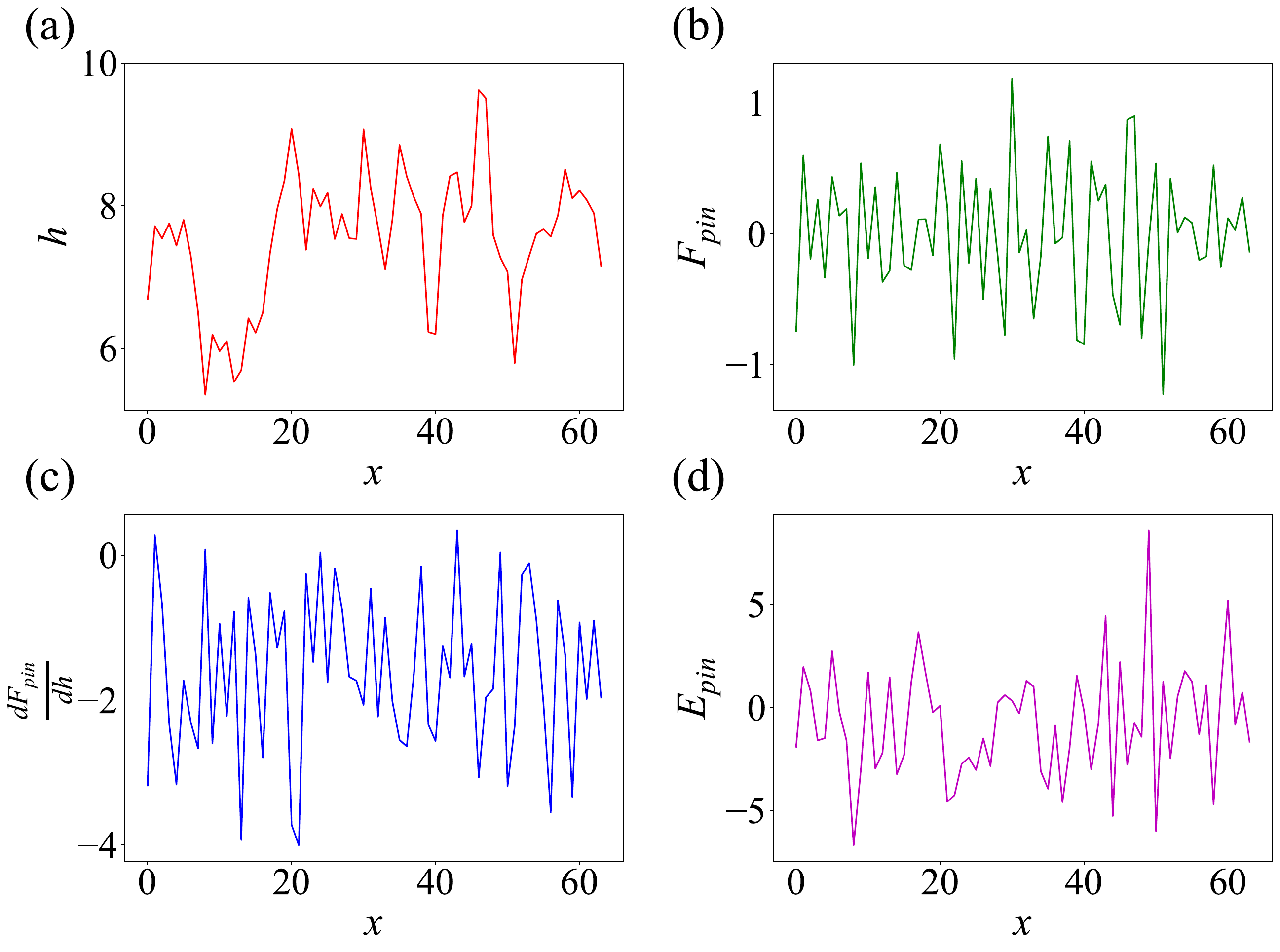}
    \caption{Examples of the 1D fields along the relaxed line configuration: (a) Line profile $h(x)$, (b) pinning force $F_\mathrm{pin}$, (c) first derivative of the pinning force field in the direction of interface motion $dF_\mathrm{pin}/dh$, and (d) the pinning potential energy $E_\mathrm{pin}$.}
    \label{fig:2}
\end{figure*}

\begin{figure*}[t!]
    \centering
    \includegraphics[width=0.6\pdfpagewidth]{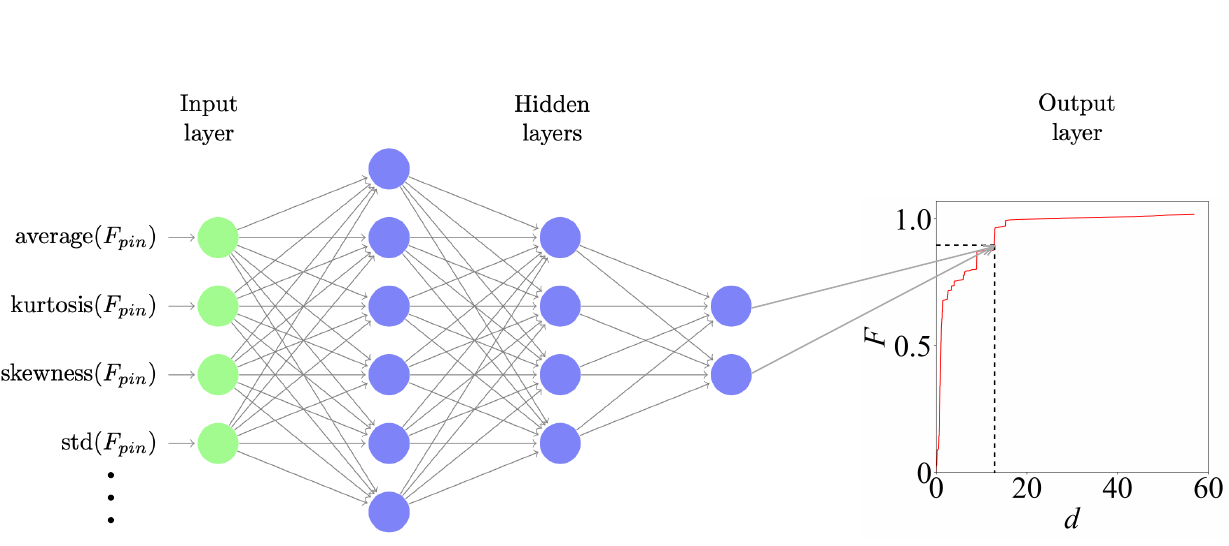}
    \caption{Schematic representation of the architecture of the NN. The statistical features of the 1D fields shown in Fig. \ref{fig:2} are fed as the input shown here as the green circles. They are subsequently passed forward to the hidden layers, whose neurons are represented by the blue circles. Finally, in the output layer the predicted value of $F(d)$ is obtained.}
    \label{fig:3}
\end{figure*}

LR is the simplest predictive model utilized in this work. As its name suggests it assumes a linear dependence of the predicted value, in this case $F(d)$, on some input features $x_i$, which can be written as
\begin{equation}
\label{eq:2}
    F(d)=\sum_{i=1}^{N_{in}}a_{i}x_{i}+b,
\end{equation}
where $N_{in}$ is the total number of the input features, while $a_i$ and $b$ are the coefficients that need to be fitted.

As the input features for LR various statistical descriptors (average, kurtosis, skewness, standard deviation, etc.) were extracted from the 1D fields along the relaxed line configuration ($h$, $F_\mathrm{pin}, \mathrm{d}F_\mathrm{pin}/\mathrm{d}h$, and the pinning energy $E_\mathrm{pin}=-\int_0^{h_i} F_\mathrm{pin}\mathrm{d}h$; examples are shown in Fig.~\ref{fig:2}). The fitting is done by using the least squares method. In addition, L1 regularization (Lasso), which adds a penalty term to the loss function that is optimized to reduce overfitting, is applied with the factor $\lambda=10^{-3}$. The whole dataset of configurations is split into the training and test set with the ratio 80:20 \%.

\begin{figure*}[t!]
    \centering
    \includegraphics[width=0.425\pdfpagewidth]{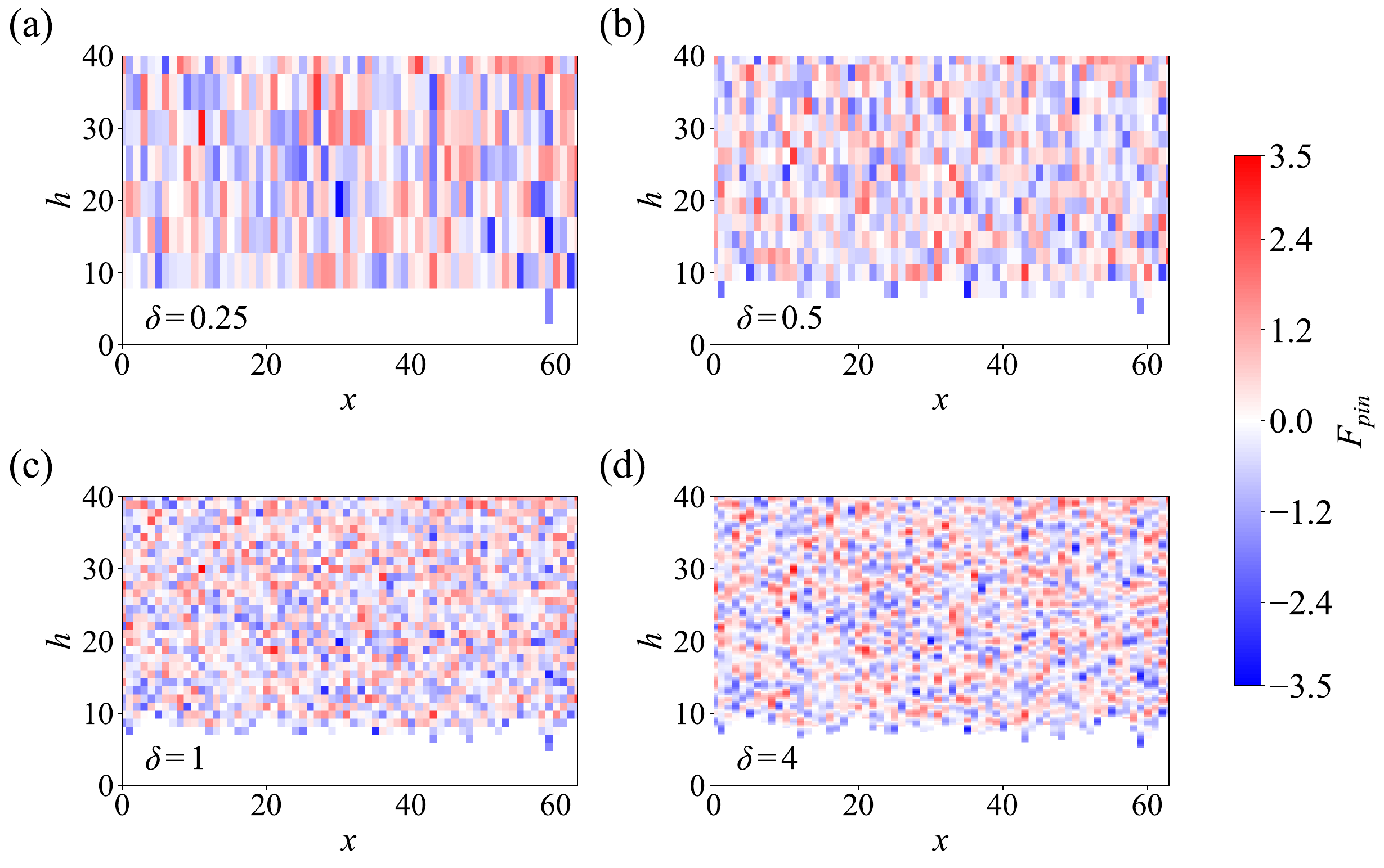}
    \caption{Examples of the two-dimensional fields of the pinning force $F_\mathrm{pin}$ above the relaxed line profile for different resolutions $\delta$ in the $h$-direction. (a) $\delta = 0.25$, (b) $\delta = 0.5$, (c) $\delta = 1$, and (d) $\delta=4$.}
    \label{fig:5}
\end{figure*}

\begin{figure*}[t!]
    \centering
    \includegraphics[width=0.425\pdfpagewidth]{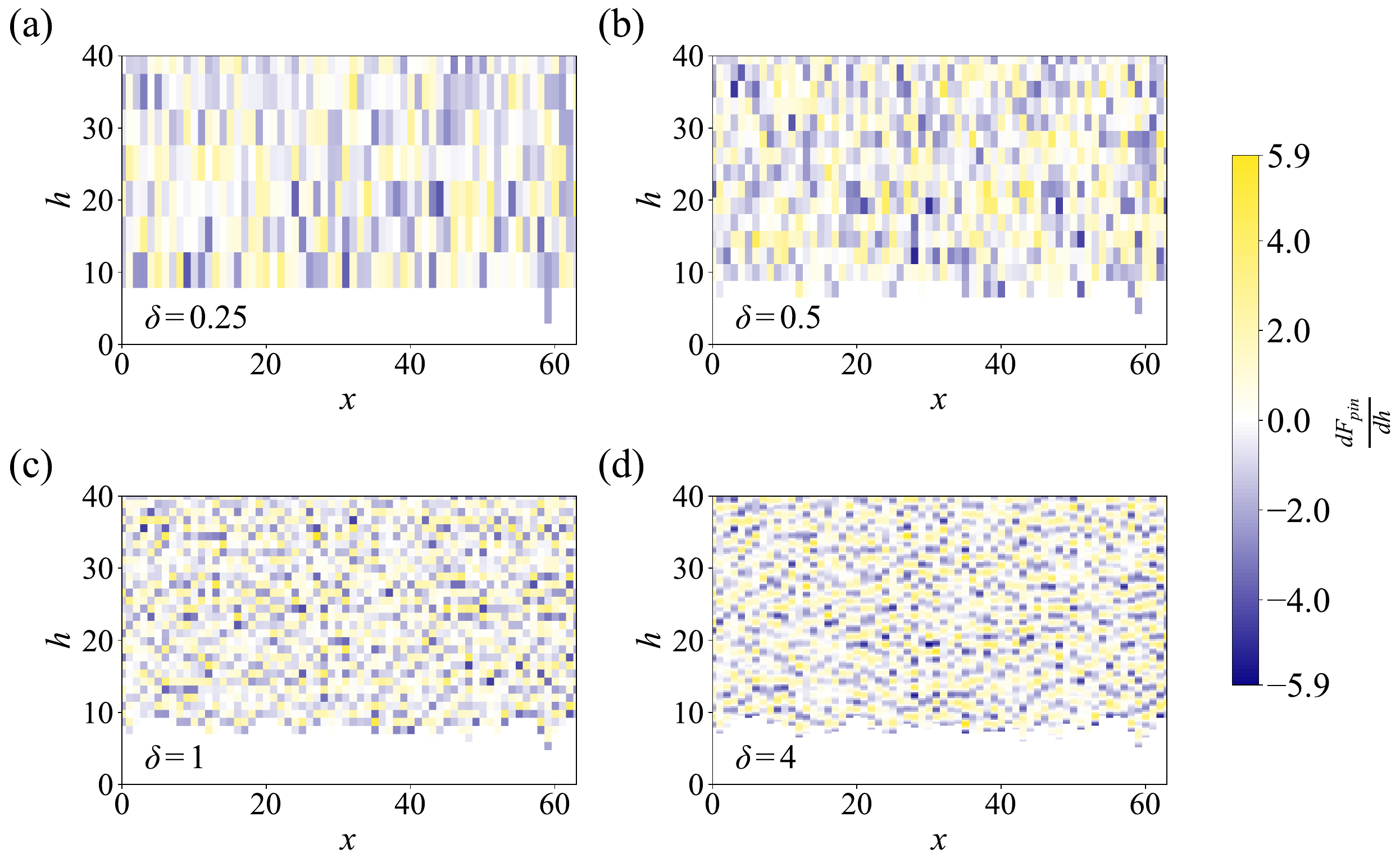}
    \caption{Examples of the two-dimensional fields of the pinning force derivative $\mathrm{d}F_\mathrm{pin}/\mathrm{d}h$ above the relaxed line profile for different resolutions $\delta$ in the $h$-direction. (a) $\delta = 0.25$, (b) $\delta = 0.5$, (c) $\delta = 1$, and (d) $\delta=4$.}
    \label{fig:5_2}
\end{figure*}

\begin{figure*}[t!]
    \centering
    \includegraphics[width=0.425\pdfpagewidth]{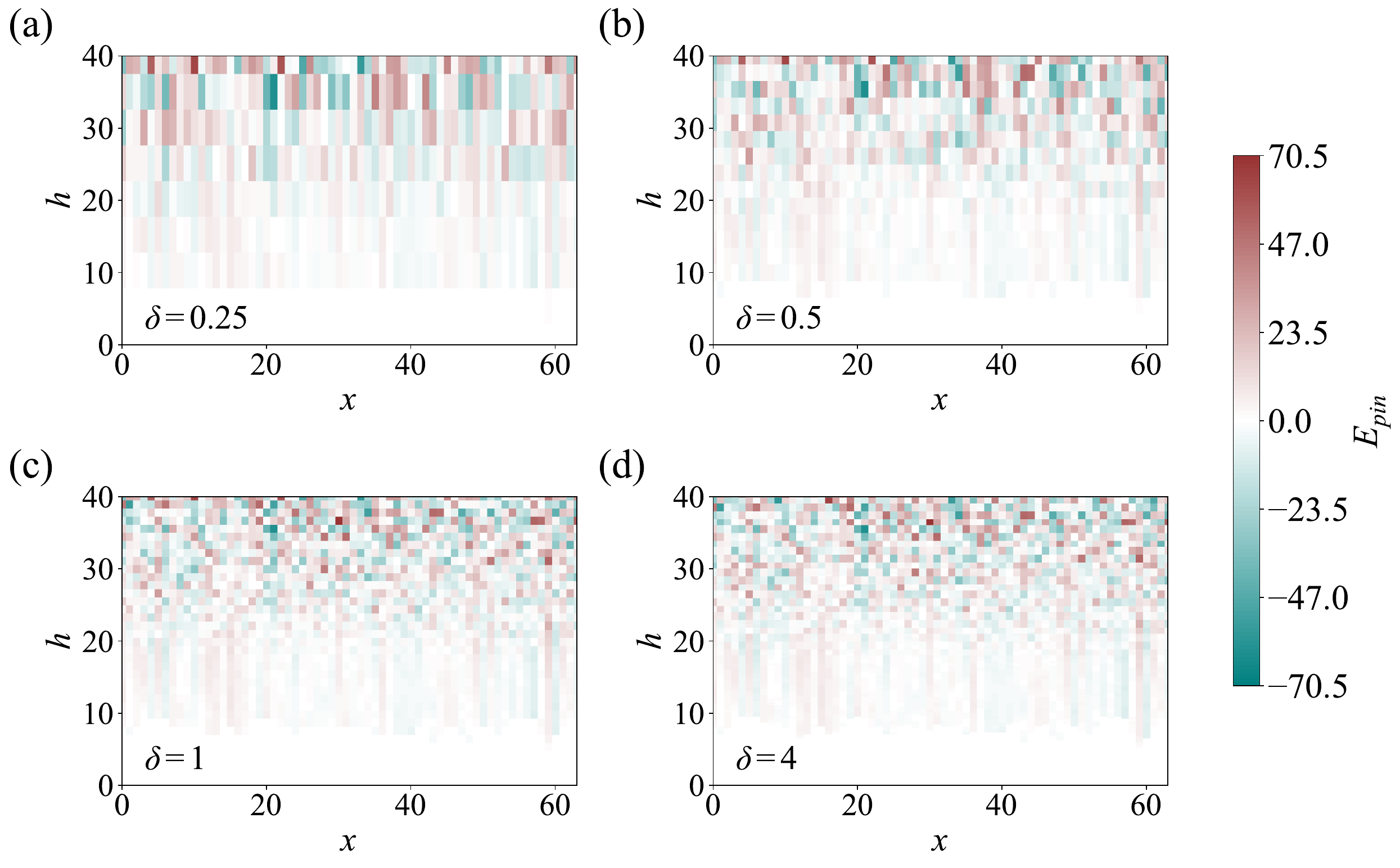}
    \caption{Examples of the two-dimensional fields of the pinning energy  $E_\mathrm{pin}$ above the relaxed line profile for different resolutions $\delta$ in the $h$-direction. (a) $\delta = 0.25$, (b) $\delta = 0.5$, (c) $\delta = 1$, and (d) $\delta=4$.}
    \label{fig:5_3}
\end{figure*}

\subsection{Fully connected neural network}

NN being a more complex model allows to infer a non-linear relation between the input and the output. The architecture of the NN is shown schematically in Fig.~\ref{fig:3}. 
It consists of an input layer to which the same statistical features of the 1D input fields as those used above for LR are fed, as well as three hidden layers, and an output layer. The features are passed forward from the input layer to the subsequent hidden layers through activation functions, which are all taken to be rectifiers (ReLU). 
Each hidden layer consists of a certain number of neurons: Starting from the first hidden layer, the hidden layers contain 64, 16 and 4 neurons, respectively. The value in the output layer, which is the predicted value of $F(d)$, is a linear function of the values of the neurons in the last hidden layer. If one denotes the value of the $m$th neuron in the $n$th layer as $y_m^n$, then
\begin{equation}
\label{eq:3}
    y_m^n=f_a^{n-1}\left(w_{0m}^{n-1}+\sum_{i=1}^{N_{n-1}}w_{im}^{n-1}y_i^{n-1}\right),
\end{equation}
where $f_a^{n-1}$ is the activation function between the $(n-1)$th and the $n$th layer, $N_{n-1}$ is the number of nodes in the $(n-1)$th layer and $w_{im}^{n-1}$ are the trainable parameters of the NN (called weights for $i\neq0$ and biases for $i=0$). In the present case, if one uses 0 as the index of the input layer, 
$y_m^0=x_m$, $N_0=N_{in}$ and $y_{0}^{4}=F(d)$.

For the training of the NN Adam optimizer is used with the learning rate $\eta=10^{-3}$. The training lasts maximally for 1000 epochs. In addition to the training and the test set also a validation set is used, such that their ratio is 80:10:10\%. The role of the validation set is to interrupt the training if its loss function (mean squared error) does not decrease for 10 consecutive epochs. L2 regularization (Ridge) is applied during the training with $\lambda=10^{-3}$.

\begin{figure}[t!]
    \centering
    \includegraphics[width=0.7\columnwidth]{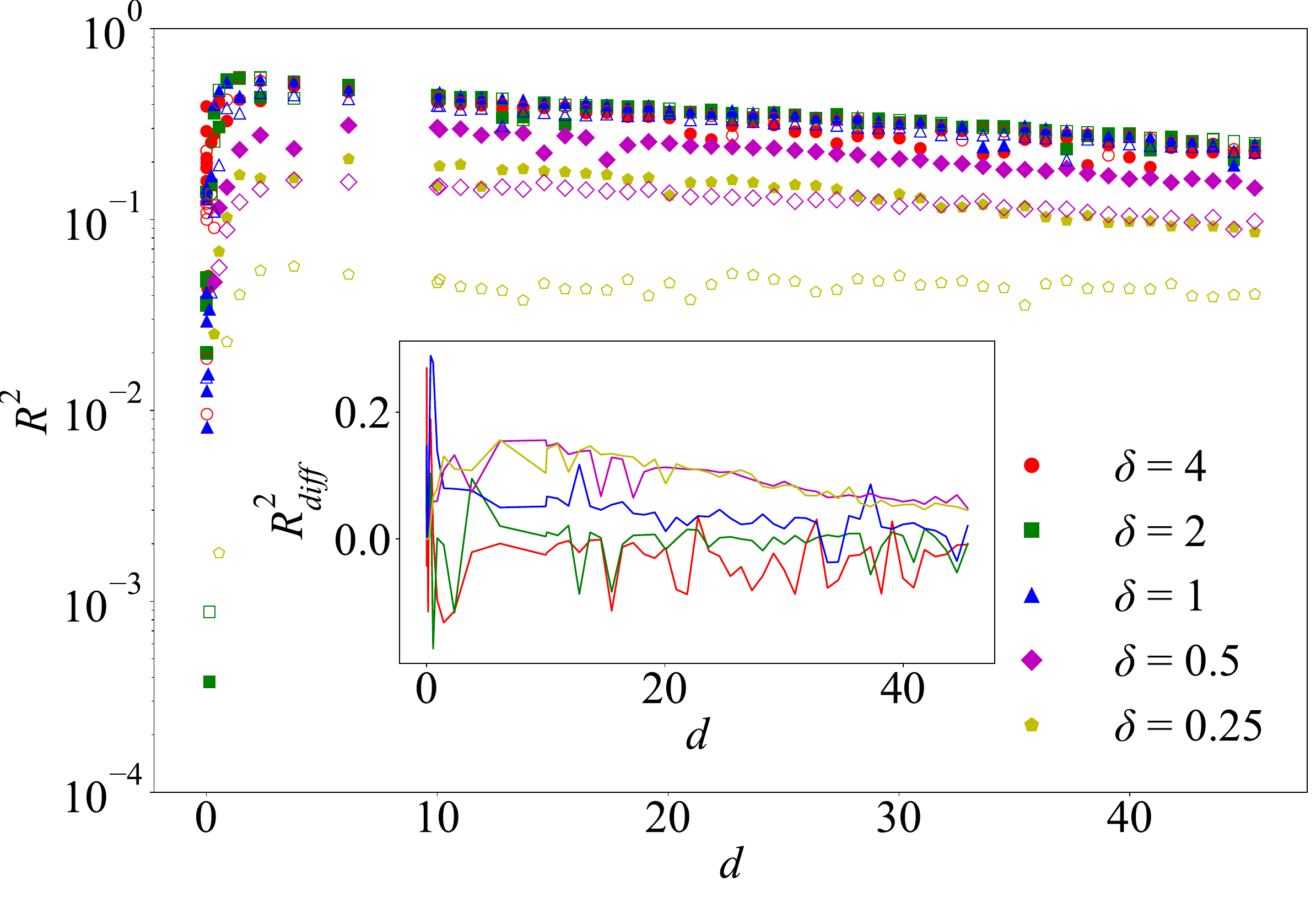}
    \caption{A comparison of $R^2$ vs $d$ for $L=64$ considering different resolutions $\delta$ when feeding the 2D CNN either all three fields ($F_\mathrm{pin}$, $\mathrm{d}F_\mathrm{pin}/\mathrm{d}h$ and $E_\mathrm{pin}$; filled symbols) or just the $F_\mathrm{pin}$ field (open symbols). The inset shows the difference $R^2_\mathrm{diff} = R^2_\mathrm{all} - R^2_{F_{\mathrm{pin}}}$ as a function of $d$ for the different $\delta$-values, with the latter indicated by the color according to the legend of the main figure.}
    \label{fig:fpin_vs_all}
\end{figure}

\subsection{Convolutional neural network}

\begin{table}[]
    \centering
    \begin{tabular}{c|c|c}
        Layer & Input shape & Number of parameters \\\hline
        PeriodicPaddingLayer\_1 & 32$\times$64$\times$3 & \\
        Conv2D\_1 & 34$\times$66$\times$3 & 448\\
        MaxPooling2D\_1 & 32$\times$64$\times$16 &\\
        PeriodicPaddingLayer\_2 & 16$\times$32$\times$16 &\\
        Conv2D\_2 & 18$\times$34$\times$16 & 2320\\
        MaxPooling2D\_2 & 16$\times$32$\times$16 &\\
        PeriodicPaddingLayer\_3 & 8$\times$16$\times$16 &\\
        Conv2D\_3 & 10$\times$18$\times$16 & 2320\\
        MaxPooling2D\_3 & 8$\times$16$\times$16 &\\
        PeriodicPaddingLayer\_4 & 4$\times$8$\times$16 &\\
        Conv2D\_4 & 6$\times$10$\times$16 & 2320\\
        MaxPooling2D\_4 & 4$\times$8$\times$16 &\\
        PeriodicPaddingLayer\_5 & 2$\times$4$\times$16 &\\
        Conv2D\_5 & 4$\times$6$\times$16 & 2320\\
        MaxPooling2D\_5 & 2$\times$4$\times$16 &\\
        Flatten & 1$\times$2$\times$16 &\\
        Dense\_1 & 32 & 136 \\
        Dense\_2 & 8 & 9
    \end{tabular}
    \caption{List of layers in 2D CNN for the 32$\times$64 input for $L=64$ and $\delta = 1$. The first dimension is equal to $\frac{L\delta}{2}$. The second dimension always corresponds to the system size $L$. All resolutions and system size combinations have the same shape in the third dimension where 3 channels represent the different fields, pinning force, pinning force derivative and pinning energy. The schematic representation of these fields can be found from the (c) subplot of Figs. \ref{fig:5}, \ref{fig:5_2} and \ref{fig:5_3}, respectively.}
    \label{tab:cnn_params}
\end{table}

Unlike LR and NN, the CNN does not utilize the manually defined features but instead takes a pixelized image of some field describing the system as the input. That image is stored in an array of the size corresponding to its resolution and subsequently processed by several repetitions of periodic padding, convolutional and pooling layers. The periodic padding layer extends the size of the array by 2 in each direction according to periodic boundary conditions. This guarantees that the array has the same size after the convolutional layer as it had before the padding layer. Convolutional layers contain convolutional filters, whose parameters act as weights and biases (analogously to the NN) for the values from the previous layer and are adjusted during the training. In each convolutional layer there are 16 filters with the kernel size of 3 and 3$\times$3 for the 1D CNN and 2D CNN, respectively. Max-pooling layers reduce the size of the array by half. Therefore, the total number of the layers is such that the size in one or both of the dimensions is eventually reduced to 1. Behind the last pooling layer a flatten layer is inserted, which converts the array to a linear one. Finally, two dense layers are added with ReLU and linear activation functions, respectively. The second to last dense layer has a size of half the number of filters followed by the last dense layer, which gives a prediction of $F(d)$ as the output. A representation of the layers, shapes and number of trainable parameters are shown in table \ref{tab:cnn_params}.

We consider two types of CNNs to predict the response of the elastic line: (i) A 1D CNN where the input fields are again given by $h$, $F_\mathrm{pin}, \mathrm{d}F_\mathrm{pin}/\mathrm{d}h$, and the pinning energy $E_\mathrm{pin}=-\int_0^{h_i} F_\mathrm{pin}\mathrm{d}h$, but instead of manually extracting statistical features from them, the full 1D fields are used as input, see Fig.~\ref{fig:4} for a schematic representation of the 1D CNN. (ii) A 2D CNN where two-dimensional maps of $F_\mathrm{pin}$, $\mathrm{d}F_\mathrm{pin}/\mathrm{d}h$ and $E_\mathrm{pin}$ above the relaxed line profile (i.e., in the direction where the line will move once $F$ is increased from zero) are used as input. Below the relaxed line profile, the maps contain zeros, and hence the interface between zeros and (mostly) non-zero values contains information of the relaxed interface configuration. The resolution of these maps is such that there are always $L$ pixels in the $x$-direction, and $\delta$ (where $\delta$ defines the resolution) pixels per unit displacement in the $h$-direction, see Figs.~\ref{fig:5}, \ref{fig:5_2}, and \ref{fig:5_3} for examples of these maps with four different resolutions $\delta$. The input shape to the first layer in the 2D CNN shown in Table \ref{tab:cnn_params} varies between resolutions and system sizes. For the default system size $L = 64$ we define a height limit $h_{lim} = 40$ in the simulation code that defines the limit of information after which we have no information of the upcoming field. For the other system sizes $L=16$, 32 and 128 the limit of information $h_{lim}=10$, 20 and 80, respectively. The example shown in Table \ref{tab:cnn_params} is thought to have a resolution of 1 since the line is initialized to a height of $L/8$ before relaxation, so forming a 32-point ($N=32$) linearly spaced array from the line's minima to $h_{lim}$ gives a resolution of $\delta = \frac{N}{h_{lim}-L/8} = \frac{32}{40-8} = \frac{32}{32} = 1$. For other resolutions of this system size, an array with 8, 16, 64 and 128 points is formed for resolutions $\delta = 0.25$, 0.5, 2, and 4, respectively. A general form for the number of points $N$ in the first dimension of the 2D CNN input layer that depends on the system and resolution can be defined as $\frac{L\delta}{2}$ where $L$ is the system size and $\delta$ the resolution. This is done to ensure that the input shape between different line configurations is the same so that it can be passed to the 2D CNN model. Also, if the whole field with resolution $\delta$ was passed to the 2D CNN, the GPU's VRAM quickly ran out especially for the larger system sizes and resolutions. A schematic representation of the 2D CNN is shown in Fig.~\ref{fig:6}.

As in the case of NN, the Adam optimizer is used for the CNN training with the learning rate $\eta=10^{-3}$ for $L=16$, $L=32$ and $L=64$, and $\eta=10^{-4}$ for $L=128$ for all the resolutions except $\delta=0.25$ where $\eta=5\cdot10^{-5}$ (a smaller learning rate was needed for the training to converge for the larger system size). The maximal number of epochs is 1000 and the ratio of the training, test and validation set is again 80:10:10\%. L2 regularization is applied with $\lambda=10^{-3}$.

\section*{Appendix B: Importance of the 2D input fields}

In our 2D CNN analysis, we used three 2D input fields: $F_\mathrm{pin}$, $\mathrm{d}F_\mathrm{pin}/\mathrm{d}h$ and 
$E_\mathrm{pin}$. Since the latter two are in principle derivable from the first, we address here the question of how much the 
predictions improve as a result of feeding the 2D CNN these three fields instead of just $F_\mathrm{pin}$. Fig.~\ref{fig:fpin_vs_all} shows the $R^2$ vs $d$ curves as predicted by the 2D CNN for the different resolutions $\delta$ in the two cases (three vs one input field). The conclusion is that including all the three fields clearly improves $R^2$ for $\delta<1$, and slightly improves it for $\delta=1$ (see the inset of Fig.~\ref{fig:fpin_vs_all}) However, for $\delta>1$, the $R^2$-values obtained when using the three input fields are indistinguishable from those obtained for $F_\mathrm{pin}$ only within statistical fluctuations. This suggests that the $E_\mathrm{pin}$ and $\mathrm{d}F_\mathrm{pin}/\mathrm{d}h$ fields are important in determining the $F(d)$ curve, but the CNN is able to figure them out from the $F_\mathrm{pin}$ data (by integration and differentiation of $F_\mathrm{pin}$, respectively) if $F_\mathrm{pin}$ is given with a high enough resolution such that no information of the non-interpolated random forces is lost.

\section*{References}

\bibliographystyle{iopart-num}
\input{main_revised.bbl}

\end{document}

%% file: main_revised.bbl
\providecommand{\noopsort}[1]{}\providecommand{\singleletter}[1]{#1}%
\providecommand{\newblock}{}